\documentclass[12pt,preprint]{emulateapj}

\usepackage{color}
\usepackage{amsmath}

\slugcomment{accepted on 20 April 2014}

\shorttitle{Dust Formation in Pop III VMSs}
\shortauthors{Nozawa et al.}

\begin{document}

\title{DUST PRODUCTION FACTORIES IN THE EARLY UNIVERSE: FORMATION OF
CARBON GRAINS \\ IN RED-SUPERGIANT WINDS OF VERY MASSIVE POPULATION III 
STARS}

\author{
Takaya Nozawa\altaffilmark{1,2},
Sung-Chul Yoon\altaffilmark{3},
Keiichi Maeda\altaffilmark{4,2},
Takashi Kozasa\altaffilmark{5} \\
Ken'ichi Nomoto\altaffilmark{2,7}, and
Norbert Langer\altaffilmark{6}
}

\altaffiltext{1}{National Astronomical Observatory of Japan,
Mitaka, Tokyo 181-8588, Japan; takaya.nozawa@nao.ac.jp}
\altaffiltext{2}{Kavli Institute for the Physics and Mathematics of 
the Universe (WPI), The University of Tokyo, Kashiwa, Chiba 277-8583, 
Japan}
\altaffiltext{3}{Department of Physics and Astronomy, 
Seoul National University, Seoul 151-747, Korea}
\altaffiltext{4}{Department of Astronomy, Kyoto University,
Sakyo-ku, Kyoto 606-8502, Japan}
\altaffiltext{5}{Department of Cosmosciences, Graduate School 
of Science, Hokkaido University, Sapporo 060-0810, Japan}
\altaffiltext{6}{Argelander-Institut f\"{u}r Astronomie der
Universit\"{a}t Bonn, Aufdem H\"{u}gel, 53121 Bonn, Germany}
\altaffiltext{7}{Hamamatsu Professor}

\begin{abstract}

We investigate the formation of dust in a stellar wind during the 
red-supergiant (RSG) phase of a very massive Population III 
star with the zero-age main sequence mass of 500 $M_\odot$.
We show that, in a carbon-rich wind with a constant velocity, 
carbon grains can form with a lognormal-like size distribution, and 
that all of the carbon available for dust formation finally condense 
into dust for wide ranges of the mass-loss rate 
((0.1--3)$\times 10^{-3}$ $M_\odot$ yr$^{-1}$)
and wind velocity (1--100 km s$^{-1}$).
We also find that the acceleration of the wind driven by newly formed 
dust suppresses the grain growth but still allows more than half of 
gas-phase carbon to be finally locked up in dust grains.
These results indicate that at most 1.7 $M_\odot$ of carbon grains 
can form in total during the RSG phase of 500 $M_\odot$ Population 
III stars.
Such a high dust yield could place very massive primordial stars as 
important sources of dust at the very early epoch of the universe 
if the initial mass function of Population III stars was top-heavy.
We also briefly discuss a new formation scenario of carbon-rich 
ultra-metal-poor stars considering the feedback from very massive 
Population III stars.

\end{abstract}


\section{Introduction}

The discoveries of huge amounts of dust grains in high-redshift 
quasars (Bertoldi et al.\ 2003; Priddey et al.\ 2003) have posed the 
fundamental problems on the origin of dust in the early universe.
At such an early epoch, core-collapse supernovae (CCSNe) arising 
from massive stars are considered to be the most promising sources 
of dust (e.g., Dwek et al.\ 2007).
On the other hand, the contribution from asymptotic giant branch 
stars evolving from intermediate-mass ($M_{\rm ZAMS} \simeq$ 3--8 
$M_\odot$) stars has also been invoked to explain a large content 
of dust in high-redshift objects
(Valiante et al.\ 2009; Dwek \& Cherchneff 2011).
What stellar mass range can mainly contribute to the dust budget in 
the early universe strongly depends on the initial mass function 
(IMF) of the stars 
(Valiante et al.\ 2011; Gall et al.\ 2011a, 2011b).

Numerical simulations of the formation of metal-free stars have shown 
that the IMF of the first generation of stars, so-called Population 
III (Pop III) stars, would be weighted towards much higher mass than 
those in the present universe 
(Bromm \& Larson 2004; Hirano et al.\ 2014).
However, a characteristic mass of Pop III stars remains to be
clarified, spanning from $\sim$40 $M_\odot$ 
(Hosokawa et al.\ 2011; Susa 2013) 
up to more than 300 $M_\odot$ 
(Omukai \& Palla 2003; Ohkubo et al.\ 2009).
In particular, Pop III stars with the masses exceeding $\sim$250 
$M_\odot$ emit numerous ionizing photons and finally collapse into 
black holes (BHs), serving as seeds of supermassive BHs.
Thus, such very massive Pop III stars would have crucial impacts on 
the reionization of the universe and dynamical evolution of galaxies.

Even though most of  very massive Pop III stars are not supposed 
to explode as supernovae (SNe), they are likely to play an important 
role in the chemical enrichment of the early universe.  
Yoon et al.\ (2012) found that non-rotating models with 
$M_\mathrm{ZAMS} > 250~M_\odot$ undergo convective dredge-up of large 
amounts of carbon and oxygen from the helium-burning core to the
hydrogen-rich envelope during the red-supergiant (RSG) phase.  
This may lead to enrichment of the surrounding medium with CNO elements 
via RSG winds. 
More importantly, such CNO-enriched RSG winds can serve as formation 
sites of dust in the early universe. 
In this Letter, we elaborate this new scenario of dust formation by
Pop III stars, using an exemplary model with $M_{\rm ZAMS} = 500$ 
$M_\odot$.
We show that C grains can form efficiently in the stellar wind with
a constant velocity for a reasonable range of mass-loss rates and wind 
velocities.
We also discuss the effect of the wind acceleration on dust formation.

\begin{deluxetable*}{cclcccc}
\tabletypesize{\scriptsize}
\tablewidth{0pt}
\tablecaption{Chemical Reactions for Formation of C Clusters 
Considered in This Paper}
\tablehead{ 
& \colhead{key molecule} & \colhead{chemical reaction} & 
\colhead{$A/10^4$K} & \colhead{$B$} & \colhead{$a_0$ (\AA)} 
& \colhead{$\sigma$ (erg cm$^{-2}$)}}
\startdata
\\
(1) Model A    &  C  &
C$_{n-1}$ + C $\rightleftharpoons$ C$_{n}$ ~~~ ($n \ge 2$)
& 8.3715  & 22.1509 &  1.281  & 1400 \\ \\
(2) Model B    & C$_2$H  &
2(C$_2$H + H) $\rightleftharpoons$ C$_{2n}$ + 2H$_2$ ~~~~~~~~~~~~~ 
($n = 2$) 
& 8.6425  & 18.9884 &  1.614  & 1400 \\
           &         &
C$_{2(n-1)}$ + C$_2$H + H $\rightleftharpoons$ 
C$_{2n}$ + H$_2$  ~~~ ($n \ge 3$) & & & \\
\enddata
\tablecomments{
The key molecule is defined as the gas species whose collisional 
frequency is the least among the reactants.
The Gibbs free energy ${\it \Delta} \mathring{g}$ for the formation 
of the condensate from reactants per key molecule is approximated by 
${\it \Delta} \mathring{g} / k T = - A/T + B$ with the numerical 
values $A$ and $B$ derived by least-squares fittings of the 
thermodynamics data (Chase et al.\ 1985).
The radius of the condensate per key molecule and the surface 
tension of bulk grains are $a_0$ and $\sigma$, respectively.}
\end{deluxetable*}

\section{THE MODEL}

For the properties of a 500 $M_\odot$ RSG, we refer to the model 
sequence m500vk00 without rotation in Yoon et al.\ (2012);
the average luminosity and effective temperature of this RSG
are $L_* = 10^{7.2}$ $L_\odot$ and $T_* = 4,440$ K, respectively, 
with a stellar radius of $R_* = 6,750$ $R_\odot$.
This very massive RSG undergoes convective dredge-up during 
helium-core burning, enriching the hydrogen envelope with a large 
amount of carbon and oxygen;
the average number fractions of the major elements in the envelope are
$A_{\rm H} = 0.701$, $A_{\rm He} = 0.294$, 
$A_{\rm C} = 3.11 \times 10^{-3}$, and 
$A_{\rm O} = 1.75 \times 10^{-3}$,
leading to a high C/O ratio (C/O  = 1.78).

\subsection{Hydrodynamic Model of the Outflowing Gas}

As the first step to assess the possibility of dust formation in a 
Pop III RSG wind, we consider a spherically symmetric gas flow with 
a constant wind velocity.
In this case, the density profile of the gas flow is given by 
\begin{eqnarray}
\rho(r) = \frac{\dot{M}}{4 \pi r^2 v_{\rm w}}
= \rho_* \left( \frac{r}{R_*} \right)^{-2},
\end{eqnarray}
where $\dot{M}$ is the mass-loss rate, $v_{\rm w}$ is the wind 
velocity, and $r$ is the distance from the center of the star.
The radial profile of the gas temperature is assumed to be
\begin{eqnarray}
T(r) = T_* \left( \frac{r}{R_*} \right)^{-\frac{1}{2}},
\end{eqnarray}
following the previous studies on dust formation in stellar winds 
(e.g., Gail et al.\ 1984).

Mass loss from a RSG was not considered in Yoon et al.\ (2012), and 
both the mass-loss rate and the wind velocity are hardly known for 
Pop III RSGs.   
Given that the underlying physics of mass-loss mechanisms is not well 
understood, modeling elaborately the mass-loss history is beyond the 
scope of this letter.
Instead, to cover various physical conditions of the mass-loss winds, 
we treat $\dot{M}$ and $v_{\rm w}$ as free parameters and examine how 
these quantities affect the formation process of dust.
In what follows, we take as fiducial values $v_{\rm w} =$ 20 km s$^{-1}$
and $\dot{M} = 3 \times 10^{-3}$ $M_\odot$ yr$^{-1}$; the latter 
corresponds to the constant mass-loss rate with which the 500 $M_\odot$ 
star loses 90\% (208 $M_\odot$) of the envelope during the last 
$7 \times 10^4$ yr of the RSG phase.\footnote{
We note that, even if there were no wind, floating-off of the loosely 
bound RSG envelope as a result of the collapse of the core into a BH 
at the end of its life may also lead to mass ejection of CNO elements 
into the interstellar medium and production of dust grains
(Zhang et al.\ 2008; Kochanek 2014).}

\subsection{Model of Dust Formation}

The calculations of dust formation are performed by applying the
formulation of non-steady-state dust formation in Nozawa \& Kozasa 
(2013).
The formulae self-consistently follow the formation of small 
clusters and the growth of grains under the consideration that the 
collisions of key molecules, defined as the gas species with the least 
collisional frequency among the reactants, control the kinetics of 
dust formation process.
The formulae enable us to evaluate the size distribution and 
condensation efficiency of newly formed grains for given temporal 
evolutions of gas temperature and density, chemical composition of 
the gas, and chemical reactions for the formation of clusters.

In a carbon-rich cool gas, all oxygen atoms are bound to carbon 
atoms to form CO molecules, and carbon atoms and/or carbon-bearing 
molecules left after the CO formation can participate in the 
formation of C clusters and grains.
The chemical equilibrium calculations along the gas flow 
(e.g., Kozasa et al.\ 1996) show that, for the physical and chemical 
conditions given above, the major carbon-bearing gas species, other 
than CO, is atomic carbon at $T \ga 1,750$ K and C$_2$H at $T \simeq$ 
1,400--1,700 K.
Thus, the formation of C clusters is expected to proceed at high
temperatures through successive attachment of carbon atoms
as given in the reaction (1) of Table 1, which we call Model A.
Independently of this, we consider another dust formation path 
involving C$_2$H, for which the possible chemical reactions are 
given under (2) of Table 1, hereafter referred to as Model B.

In the calculations, we assume that a fraction $f_{\rm C}$ of the 
carbon that is not locked up in CO molecules exists as carbon 
atoms in Model A and as C$_2$H molecules in Model B.
Since $f_{\rm C}$ linearly changes the number density of carbon 
available for dust formation, decreasing $f_{\rm C}$ is identical with 
reducing the mass-loss rate or the C/O ratio in the envelope by the 
same factor.
The time evolutions of the gas density and temperature are
calculated by substituting $r = R_* + v_{\rm w} t$ into Equations 
(1) and (2).
The sticking probability of gas species is assumed to be unity, and C 
clusters that contain more carbon atoms than $n_* = 100$ are treated 
as bulk grains. 
We refer the readers to Nozawa \& Kozasa (2013) for the formulation
of dust formation process and the detailed prescription of the 
calculations.

\begin{figure}
\epsscale{1.1}
\plotone{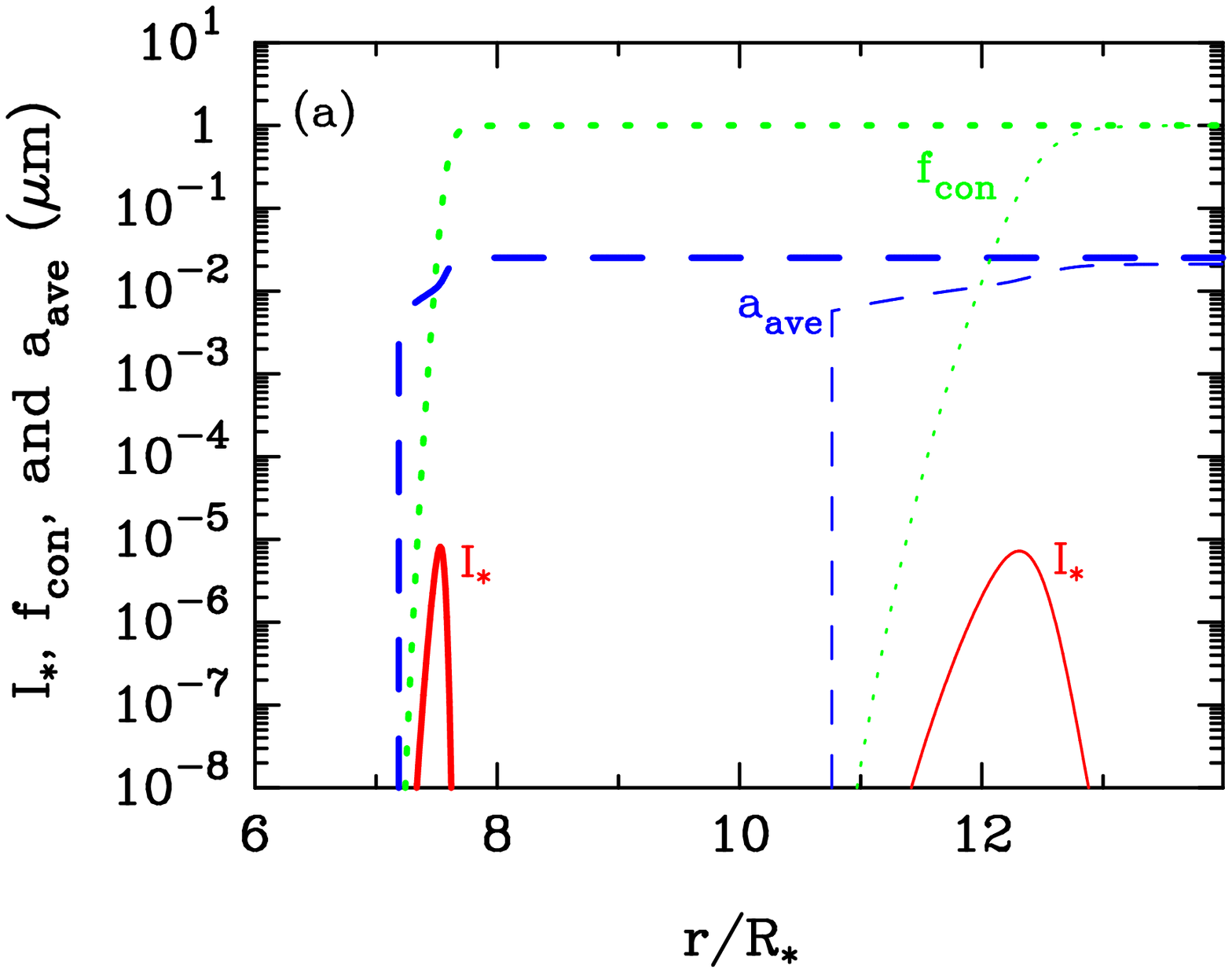}
\vspace{0.2 cm}
\plotone{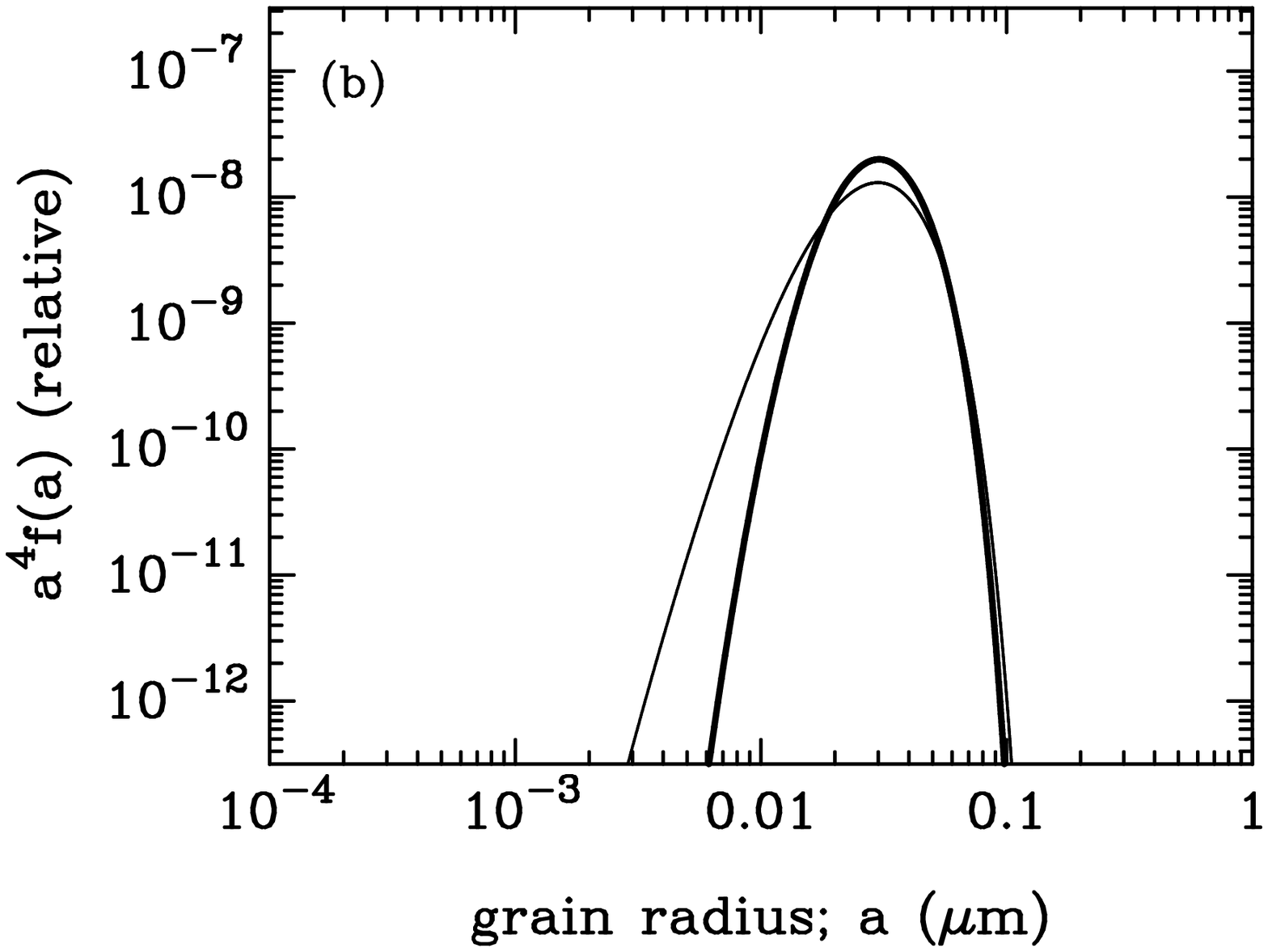}
\caption{
(a) Formation rate of seed clusters with $n_* = 100$, divided by the 
nominal concentration of the key molecules ($I_*$, solid), 
condensation efficiency ($f_{\rm con}$, dotted), and average grain
radius ($a_{\rm ave}$, dashed) as a function of distance from the 
center of the star ($r/R_*$), and (b) final size distribution spectrum 
by mass $a^4 f(a)$ of newly formed C grains for a mass-loss rate 
$\dot{M} = 3 \times 10^{-3}$ $M_\odot$ yr$^{-1}$, a wind velocity 
$v_{\rm w} = 20$ km s$^{-1}$, and $f_{\rm C} = 1$.
The thick lines are for the Model A where the chemical reaction (1)
in Table 1 is considered for the formation of clusters, while the thin 
lines are for the Model B with the chemical reactions (2).
\label{fig1}}
\end{figure}

\section{RESULTS OF DUST FORMATION CALCULATIONS}

Figure 1 shows the results of the calculations for the fiducial case
with $f_{\rm C} \dot{M} = 3 \times 10^{-3}$ $M_\odot$ yr$^{-1}$ and 
$v_{\rm w} = 20$ km s$^{-1}$;
Figure 1(a) plots the formation rate of seed clusters with $n_* = 100$ 
divided by the concentration of key species without depletion 
due to cluster/grain formation ($I_*$), condensation efficiency 
($f_{\rm con}$), and average grain radius ($a_{\rm ave}$) as a 
function of distance from the center of the star ($r/R_*$).
Here, the condensation efficiency $f_{\rm con}(t)$ is defined as 
the fraction of free carbon atoms that are locked up in grains.
In Model A (thick lines) and Model B (thin lines), dust grains start 
to form at 7.2 $R_*$ and 10.8 $R_*$, respectively, with $I_*$ being 
peaked around 7.5 $R_*$ and 12.3 $R_*$.
In both of the models, the final condensation efficiency 
$f_{{\rm con}, \infty} = f_{\rm con} (t \rightarrow \infty)$ 
is unity.

The final average grain radius is only a little higher 
for Model A ($a_{{\rm ave},\infty} = 0.025$ $\mu$m) than 
for Model B ($a_{{\rm ave},\infty} = 0.021$ $\mu$m);
for Model B, the concentration of key molecules at the time of 
dust formation is lower than for Model A by a factor of $\simeq$5,
but the resulting decrease in the formation rate of seed clusters 
is compensated with the decrease in the growth rate.
As a result, the final average radius is similar in both Model A 
and Model B, although lower rates (longer timescales) of both 
processes for Model B lead to a broader lognormal-like size 
distribution of grains, as seen from Figure 1(b).
These results demonstrate that the final condensation efficiency and 
average grain radius are almost independent of the chemical reactions 
for the formation of C clusters in the context of this study.

\begin{figure}
\epsscale{1.1}
\plotone{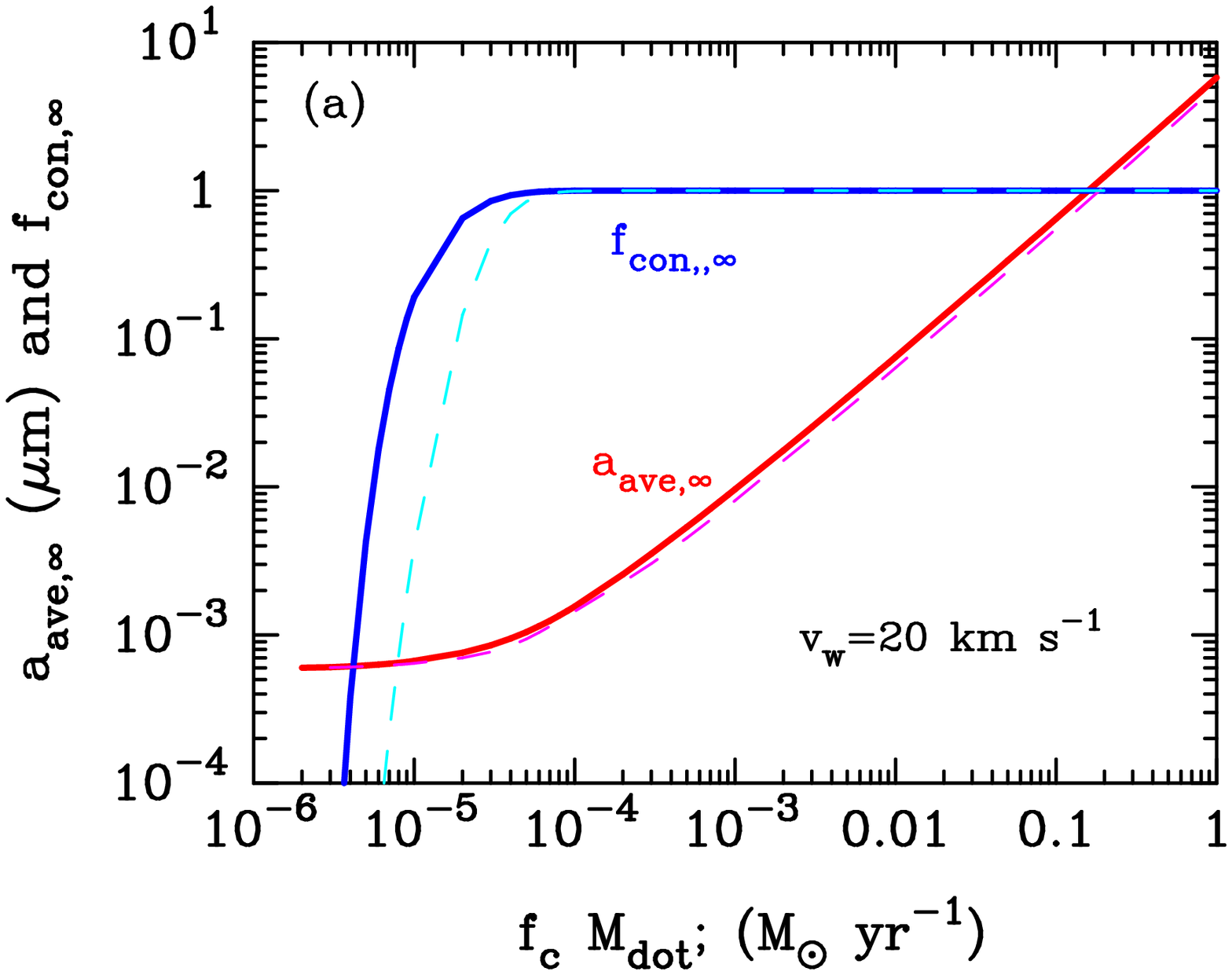}
\vspace{0.2 cm}
\plotone{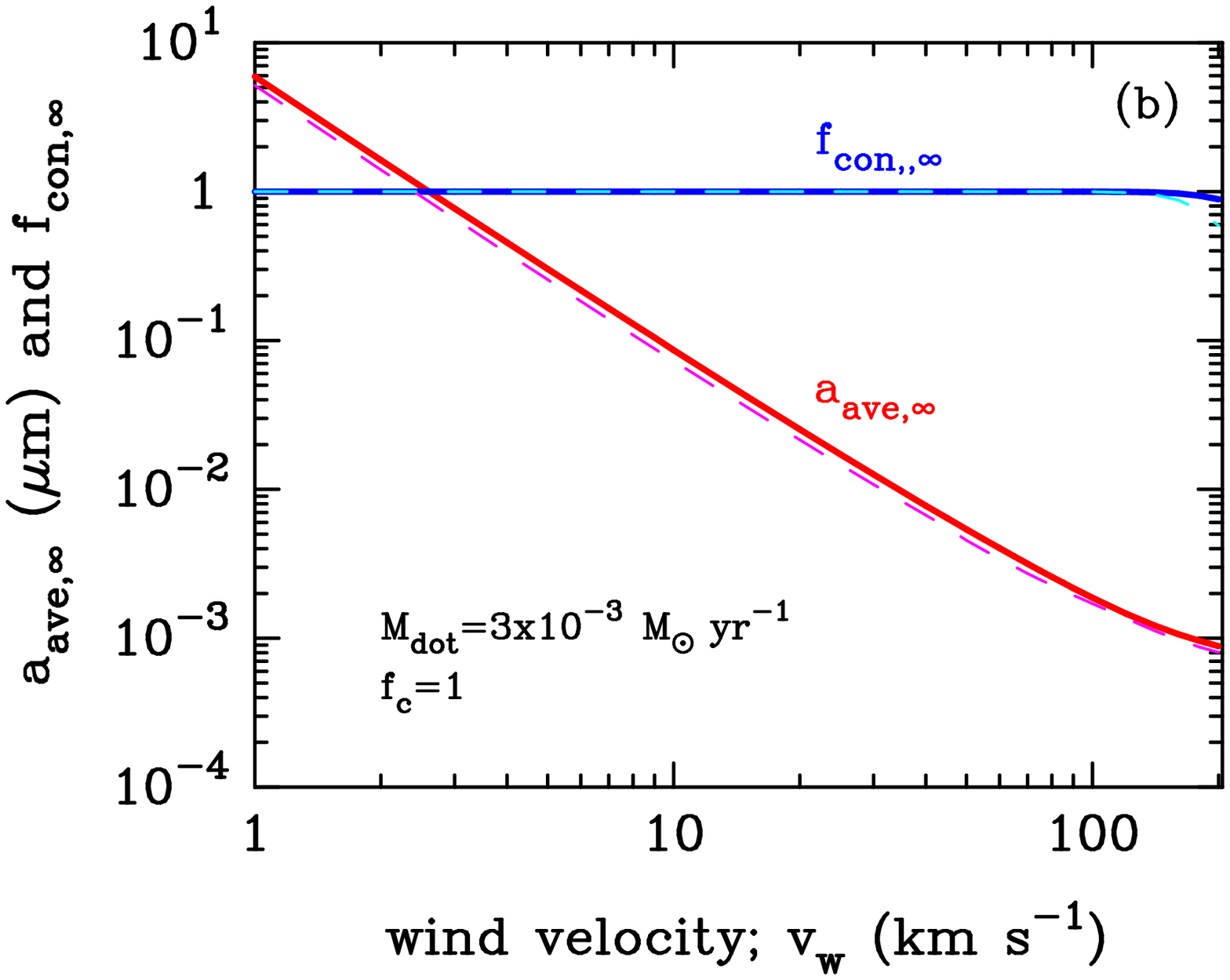}
\caption{
Final average radius $a_{{\rm ave},\infty}$ and final condensation 
efficiency $f_{{\rm con}, \infty}$ of C grains formed in the outflowing 
gas;
(a) as a function of product $f_{\rm C} \dot{M}$ for 
$v_{\rm w} = 20$ km s$^{-1}$, and
(b) as a function of $v_{\rm w}$ for 
$f_{\rm C} \dot{M} = 3 \times 10^{-3}$ $M_\odot$ yr$^{-1}$.
The thick solid lines are for Model A, while the thin dashed lines
are for Model B.
\label{fig1}}
\end{figure}

Figure 2(a) indicates the final condensation efficiency and final 
average radius of newly formed C grains as a function of product 
$f_{\rm C} \dot{M}$ for $v_{\rm w} =$ 20 km s$^{-1}$.
For both Model A and Model B, $f_{{\rm con}, \infty} = 1$ at 
$f_{\rm C} \dot{M} \ga 10^{-4}$ $M_\odot$ yr$^{-1}$, 
where the average grain radius scales as 
$a_{{\rm ave}, \infty} \propto (f_{\rm C} \dot{M})^{0.88}$.
On the other hand, $a_{{\rm ave}, \infty}$ is more sensitive to the 
wind velocity, as seen from Figure 2(b);
$a_{{\rm ave}, \infty}$ is smaller for a higher $v_{\rm w}$ and 
scales as $a_{{\rm ave}, \infty} \propto v_{\rm w}^{-1.75}$ 
for $v_{\rm w} =$ 1--100 km s$^{-1}$.
The increase in $v_{\rm w}$ leads to a lower gas density for a fixed 
$\dot{M}$  and causes more rapid cooling 
of the gas, both of which favor producing a number of smaller grains.

The results of the calculations show that the final condensation 
efficiency of C grains is unity if the following condition is met;
\begin{eqnarray}
\left( \frac{f_{\rm C} \dot{M}}{3 \times 10^{-3} ~
M_\odot ~ {\rm yr}^{-1}} \right)
\left( \frac{v_{\rm w}}{20 ~ {\rm km} ~ {\rm s}^{-1}}
\right)^{-2}  \ga 0.04.
\end{eqnarray}
Thus, as an example, 
for $v_{\rm w} =$ 20 km s$^{-1}$ and $f_{\rm C} = 1$, the total mass 
of C grains produced over the lifetime of the RSG with 
$M_{\rm ZAMS} = 500$ $M_\odot$ is estimated as
$M_{\rm dust}/M_\odot = 1.7$
($\dot{M} / 3 \times 10^{-3}$ $M_\odot$ yr$^{-1}$) 
for $1 \times 10^{-4}$ $M_\odot$ yr$^{-1}$
$\le \dot{M} \le$ $3 \times 10^{-3}$ $M_\odot$ yr$^{-1}$.
It should be emphasized here that these newly formed 
grains could not be destroyed by the blast wave resulting from a SN 
explosion because such a very massive Pop III star would finally collapse 
into a BH (Heger \& Woosley 2002; Yoon et al.\ 2012, but see also 
Ohkubo et al.\ 2006).
The ratio of dust mass to the initial stellar mass 
($X_{\rm VMS} = M_{\rm dust} / M_{\rm ZAMS} \le 3.4 \times 10^{-3}$)
and the dust-to-metal ratio 
($M_{\rm dust} / M_{\rm metal} \le 0.24$) are in the ranges of those 
supplied by Pop III CCSNe, for which 
$X_{\rm CCSN} =$ (0.1--30)$\times 10^{-3}$ and 
$M_{\rm dust} / M_{\rm metal} =$ 0.01--0.25, depending on the 
destruction efficiency of newly formed dust by the reverse shocks 
(Bianchi \& Schneider 2007; Nozawa et al.\ 2007).
This implies that, if very massive Pop III stars had really formed, 
they could be one of rapid and efficient sources of C grains in 
the early universe.

\begin{figure}
\epsscale{1.086}
\plotone{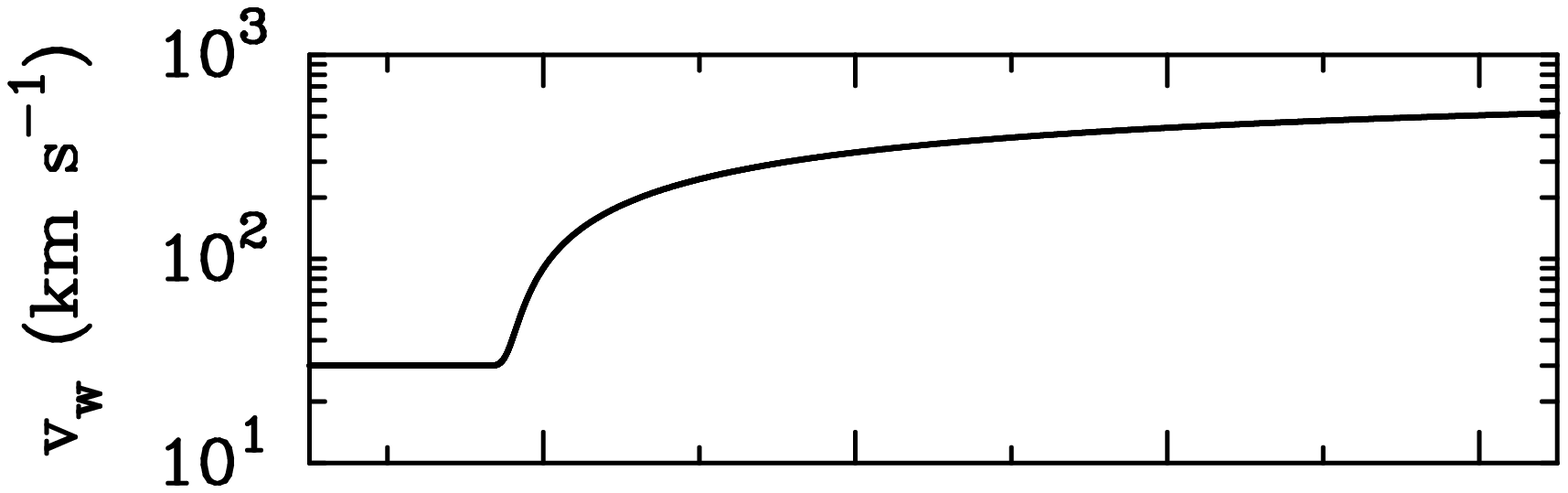}
\epsscale{1.1}
\plotone{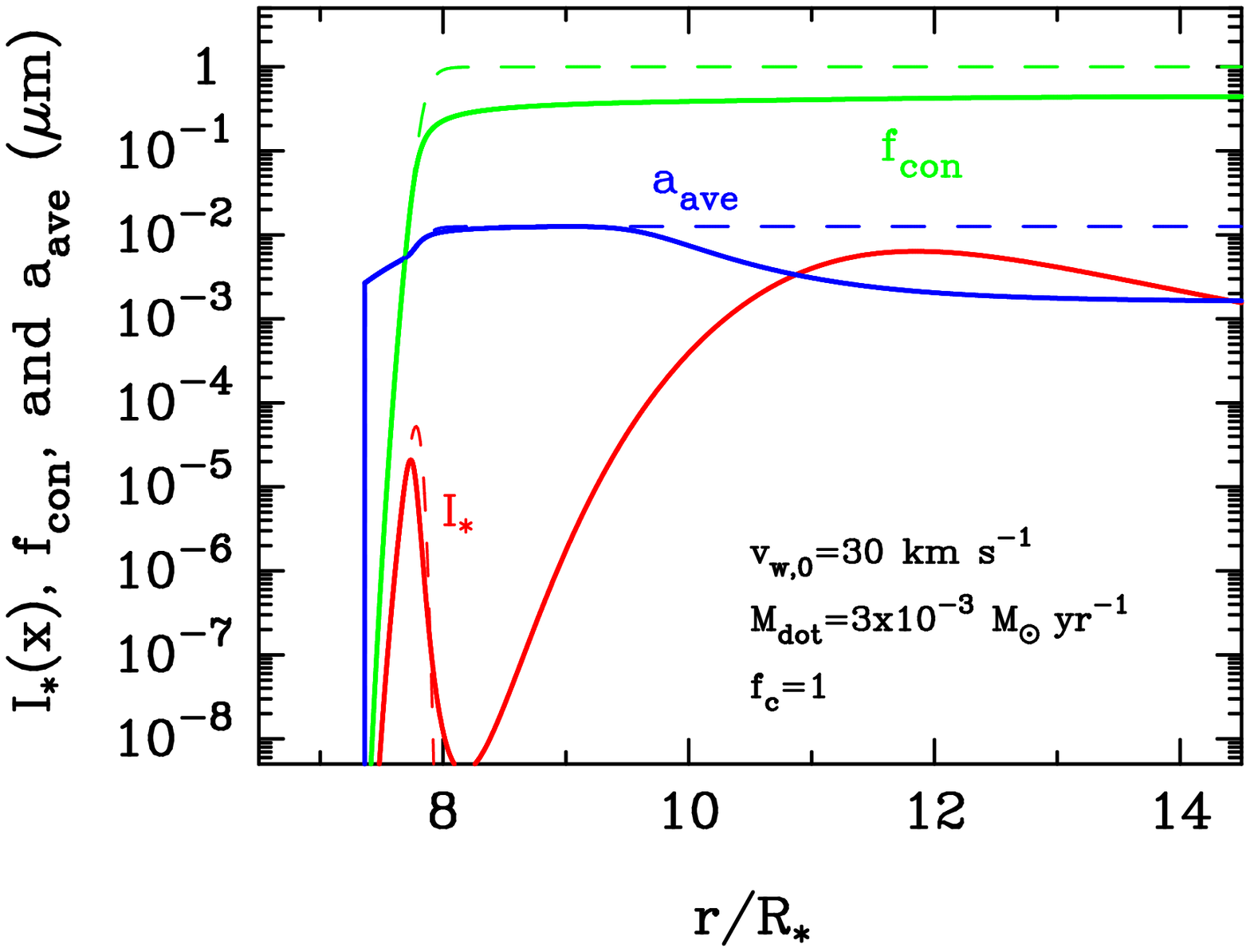}
\caption{
Wind velocity $v_{\rm  w}$ (upper panel), and formation rate of 
seed clusters $I_*$, condensation efficiency $f_{\rm con}$, and 
average grain radius $a_{\rm ave}$ (lower panel) as a function of 
$r/R_*$ for Model A with the wind acceleration.
The initial wind velocity and mass-loss rate are set to be
$v_{{\rm  w}, 0} = 30$ km s$^{-1}$ and $\dot{M} = 3 \times 10^{-3}$ 
$M_\odot$ yr$^{-1}$, respectively.
The dashed lines in the lower panel are the results without the 
wind acceleration.\\
\label{fig3}}
\end{figure}

\section{Effects of Wind Acceleration on Dust Formation}

In the previous section, we have considered the formation of dust 
in stellar winds with constant velocities.
However, the radiation pressure acting on newly formed grains will 
drive the wind to higher outflow velocities, which may suppress the 
growth of the dust grains.
Here, according to Ferarroti \& Gail (2006), we examine the effect 
of the wind acceleration on dust formation by solving the following 
simplified momentum equation:
\begin{eqnarray}
v_{\rm w} \frac{d v_{\rm w}}{dr} = - \frac{G M_*}{r^2}
\left[ 1 - \frac{L_* \langle \kappa_{\rm ext}(T)\rangle}
{4 \pi c G M_*} D \right],
\end{eqnarray}
where $G$ is the gravitational constant, $c$ is the light speed, 
$D$ is the dust-to-gas mass ratio, and 
$\langle \kappa_{\rm ext}(T) \rangle$ is the 
Planck-averaged mass extinction coefficient of C grains.
We take $\langle \kappa_{\rm ext} \rangle = 2.1 \times 10^4$ cm$^2$ 
g$^{-1}$ (for $T=4,440$ K, Zubko et al.\ 1996) and $M_* = 400$ 
$M_\odot$ as a representative value.

Figure 3 shows the acceleration of the wind and the formation process 
of dust for the initial outflow velocity of $v_{{\rm w},0} =$ 30 km 
s$^{-1}$ and the mass-loss rate of $\dot{M} = 3 \times 10^{-3}$ 
$M_\odot$ yr $^{-1}$ in Model A.
Because of the high stellar luminosity, the wind is rapidly 
accelerated to $\ge$100 km s$^{-1}$ once $f_{\rm con}$ is above 
$\simeq 2 \times 10^{-3}$.
The resulting rapid dilution of the gas largely decreases both the 
growth rate of grains and the formation rate of seed clusters but 
still allows the dust grains to grow slowly.
Furthermore, the expansion of the gas reduces the gas temperature, 
so very small grains continue condensing from carbon atoms that were 
not locked up in dust grains, as seen from the later increase in $I_*$.
As a consequence, $a_{{\rm ave}, \infty}$ becomes very small 
as a whole, and $f_{{\rm con}, \infty}$ increases to 0.45.

\begin{figure}
\epsscale{1.1}
\plotone{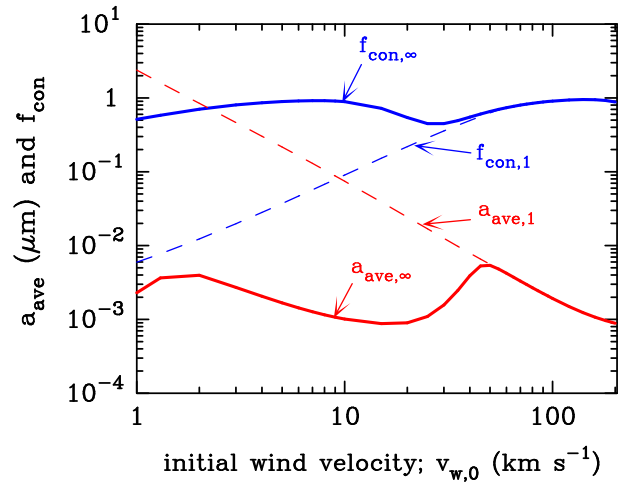}
\caption{
The solid lines show the dependence of $a_{{\rm ave},\infty}$ and 
$f_{{\rm con}, \infty}$ on $v_{{\rm w} ,0}$ in the case with the 
wind acceleration for Model A with 
$f_{\rm C} \dot{M} = 3 \times 10^{-3}$ $M_\odot$ yr$^{-1}$.
The dashed lines plot the average radius $a_{{\rm ave},1}$ and 
condensation efficiency $f_{{\rm con},1}$ just before the grain 
growth is depressed by the wind acceleration.
\label{fig1}}
\end{figure}

The dashed lines in Figure 4 plot the average radius 
($a_{{\rm ave},1}$) and condensation efficiency ($f_{{\rm con},1}$)
at the time ($t=t_1$) just before the grain growth is suppressed 
by the wind acceleration, as a function of $v_{{\rm w},0}$ for 
Model A with 
$f_{\rm C} \dot{M} = 3 \times 10^{-3}$ $M_\odot$ yr$^{-1}$.
For a lower $v_{{\rm w},0}$ with which dust grains form in the 
region closer to the star, the gas outflow is more efficiently 
accelerated, and the condensation efficiency of (large) grains formed 
before the wind acceleration is smaller.
Nevertheless, the formation of small grains at later phases, as well 
as the gradual growth of large grains, enhances $f_{{\rm con}, \infty}$ 
up to the range of 0.45--0.95 with the very small 
$a_{{\rm ave}, \infty}$ (see the solid lines in Figure 4).
Thus, the wind acceleration influences the size distribution of dust
but is not likely to affect the final condensation efficiency 
significantly.

Note that, in these calculations, we consider the acceleration of the 
winds by assuming the position coupling between the dust and the gas.
In reality, dust grains pushed by the radiation pressure move
outwards relative to the gas, then the drag force between them  
drives the acceleration of the outflowing gas.
Thus, the wind acceleration must be less efficient than that in this 
study.
On the other hand, the high-velocity motion of dust relative to the 
gas can cause the erosion of dust by sputtering 
(Tielens et al.\ 1994; Nozawa et al.\ 2006).
In particular, dust grains are accelerated above 100 km s$^{-1}$ in 
the present case, so the processing of dust by sputtering is expected 
to have considerable impacts on the final condensation efficiency.
These processes will be explored in the future work.

\section{CONCLUSION AND DISCUSSION}

We have investigated the formation of C grains in a mass-loss wind
of a Pop III RSG with $M_{\rm ZAMS} = 500$ $M_\odot$.
We find that, in a stellar wind with a constant velocity, the 
condensation efficiency of C grains is unity under the condition in 
Equation (3), and that at most 1.7 $M_\odot$ of C grains can be 
produced during the lifetimes of Pop III RSGs.
We also find that the wind acceleration caused by newly formed dust
can change the final size distribution of the dust, but still leads 
to the high final condensation efficiency 
($f_{{\rm con}, \infty} \ga 0.5$).
Such dust masses would be high enough to have an impact on the dust 
enrichment history in the early universe if the IMF of Pop III stars 
was top-heavy.

Recent sophisticated simulations of the first star formation 
(Hirano et al.\ 2014) have suggested that the number of very massive 
stars (VMSs) with $M_{\rm ZAMS} \ga 250$ $M_\odot$ ($N_{\rm VMS}$) is 
likely to be as large as that of massive stars exploding as CCSNe 
($N_{\rm CCSN}$).
If this is true and if all of the VMSs lead to 
$X_{\rm VMS} = 3.4 \times 10^{-3}$, the contribution of the 
interstellar dust from VMSs is comparable with, or even 
higher ($N_{\rm VMS} X_{\rm VMS}/N_{\rm CCSN} X_{\rm CCSN} \ga 1$) 
than that from CCSNe in the case that the destruction of dust by the 
reverse shock is efficient 
($X_{\rm CCSN} \la 1.0 \times 10^{-3}$).\footnote{
For pair-instability SNe occurring from stars with 
$M_{\rm ZAMS} \simeq$ 130--250 $M_\odot$, 
$X_{\rm PISN} \la$ 0.05 and 
$M_{\rm dust} / M_{\rm metal} \la$ 0.15, depending on the 
destruction efficiency of dust by the reverse shocks 
(Nozawa et al.\ 2007).
We also note that pair-instability SNe might be inefficient sources of 
C grains (Nozawa et al.\ 2003).}
Thus, very massive Pop III stars could be potentially dominant sources 
of dust grains at very early times of the universe.

Our results also have important indications on the formation scenario
of carbon-rich ultra-metal-poor (UMP) stars with [Fe/H] $< -4$, which
would record the chemical imprints of Pop III stars 
(Beers \& Christlieb 2005).
The formation of such low-mass metal-poor stars is considered to be
triggered through the cooling of gas by dust ejected from Pop III SNe 
(Schneider et al.\ 2012a, 2012b; Chiaki et al.\ 2013).
Ji et al.\ (2014) suggested that the formation of carbon-rich UMP 
stars relies on the cooling by fine structure lines of C and O atoms, 
assuming that the first SNe produced no C grain.

Here we propose another possible channel for the formation of 
carbon-rich UMP stars.
As shown in this study, very massive Pop III RSGs are efficient 
sources of C grains as well as CNO elements.
Thus, in the gas clouds enriched by these Pop III RSGs, C grains 
enable the formation of low-mass stars whose chemical compositions
are highly enhanced in carbon and oxygen.
As the investigated 500 $M_\odot$ model undergoes mild hot-bottom 
burning, some nitrogen is also produced, giving rise to 
[N/C] = $-4.2$ to $-1.3$ depending on the assumed mass-loss history, 
where observations of carbon-rich UMP stars indicate [N/C] $\ge -1.7$
(Christlieb et al.\ 2002; Norris et al.\ 2007; Frebel et al.\ 2008).
From our zero-metallicity model, we do not predict the presence of 
any heavier metals.
Further observations and more quantitative theoretical studies
are needed to show whether any UMP stars have formed through our
scenario.

\acknowledgments

We are grateful to the anonymous referees for critical comments.
This research has been supported by World Premier International 
Research Center Initiative (WPI Initiative), MEXT, Japan, and by the 
Grant-in-Aid for JSPS Scientific Research (22684004, 23224004, 
23540262, 26400222).

\end{document}